\documentclass[aps,prl,twocolumn,groupedaddress,showpacs]{revtex4-1}

\usepackage{graphicx}
\usepackage{dcolumn}
\usepackage{bm}
\usepackage{braket}

\usepackage{amsmath,amsthm,amscd,amssymb,amsfonts, amsbsy}
\usepackage{latexsym}
\usepackage{txfonts}

\usepackage{pgf}
\usepackage{color}

\begin{document}
\title{Dual hidden landscapes in Anderson localization on discrete lattices}
 
\author{M.L.Lyra}
\affiliation{Instituto de F\'{\i}sica, Universidade Federal
de Alagoas, 57072-970 Macei\'o, AL, Brazil}
\altaffiliation{Physique de la Mati\`ere Condens\'ee, Ecole Polytechnique, CNRS, Palaiseau, France}
\author{S. Mayboroda}
\affiliation{School of Mathematics, University of Minnesota, Minneapolis, United States of America}
\author{M. Filoche}
\affiliation{Physique de la Mati\`ere Condens\'ee, Ecole Polytechnique, CNRS, Palaiseau, France}
\altaffiliation{CMLA, ENS Cachan, CNRS, UniverSud, Cachan, France}

\date{\today}

\begin{abstract}
The localization subregions of stationary waves in continuous disordered media have been recently demonstrated to be governed by a hidden landscape that is the solution of a Dirichlet problem expressed with the wave operator. In this theory, the strength of Anderson localization confinement is determined by this landscape, and continuously decreases as the energy increases. However, this picture has to be changed in discrete lattices in which the eigenmodes close to the edge of the first Brillouin zone are as localized as the low energy ones. Here we show that in a 1D discrete lattice, the localization of low and high energy modes is governed by two different landscapes, the high energy landscape being the solution of a dual Dirichlet problem deduced from the low energy one using the symmetries of the Hamiltonian. We illustrate this feature using the one-dimensional tight-binding Hamiltonian with random on-site potentials as a prototype model. Moreover we show that, besides unveiling the subregions of Anderson localization, these dual landscapes also provide an accurate overall estimate of the localization length over the energy spectrum, especially in the weak disorder regime.
\end{abstract}

\maketitle

In Anderson localization~\cite{Anderson1958,Lagendijk2009}, electronic states are exponentially localized in a disordered alloy despite the absence of classical confinement, this localization being explained as originating from the destructive interference of waves reflected in the random atomic potential. Despite numerous theoretical advances, such as the prediction by the scaling theory of the lower critical dimension of the Anderson transition~\cite{Abrahams1979}, there was until recently no general formalism capable to accurately pinpoint the spatial location of the localized modes for any given potential, nor to predict the exact energy at which delocalized modes would begin to form.

Recently, a new theory has been proposed, unveiling a direct relationship between any specific realization of the random potential and the corresponding location of localized states~\cite{Filoche2012}. It has been demonstrated that the boundaries of the localization regions, which cannot be deduced by directly looking at the bare random potential, can be accurately retrieved as the valleys lines of a ``hidden landscape'' $u(\bf{x})$ which is the solution of a Dirichlet problem with uniform right hand side for the same Hamiltonian. This theory also predicts the progressive vanishing of these boundaries at higher energy, hence the transition from localized states to increasingly delocalized states.

This new approach to Anderson localization was proved in continuous media. In this paper, we show that not only the exact same theory can be extended to the case of a tight binding Hamiltonian defined on a discrete lattice, but also that, contrary to the continuous case, two different types of localization occur here. First, localization of low energy states (of characteristic wavelength much larger than the lattice spacing) can be predicted using a discrete analog of the landscape~$u(\bf{x})$ defined in the continuous situation. Secondly, the discreteness of the lattice also triggers a strong localization of states of typical wavelength of the order of the lattice spacing~\cite{John1987,Quang1997,Deych2003} (which would correspond to the top of the band for a periodic potential). We show here that this localization can also be studied in the framework of the landscape theory, with a different operator than the original Hamiltonian and, respectively a different landscape.

Let us first recall the essential aspects of the theory developed in~\citep{Filoche2012}. In a continuum space, the eigenstates for one particle of mass $m$ in the presence of a quenched disordered potential~$V({\bf x})$ are solutions of the time-independent Schr\"odinger equation
\begin{equation}
\label{eq:schrodinger}
[-\Delta + V({\bf x})]~\Psi({\bf x})~=~E~\Psi({\bf x}) ,
\end{equation}
where units of $\hbar^2/2m$ were considered. The only constraint we impose here on the potential is that it has to be non negative everywhere: $V({\bf x})\ge 0$. This condition can be easily fulfilled by shifting the potential without changing the eigenstates. The new approach allows us to infer several aspects of the eigenstates localization based on a single hidden landscape~$u({\bf x})$ which is actually the solution of the corresponding Dirichlet problem 
\begin{equation}
 [-\Delta + V({\bf x})]~u({\bf x})~=~1,
\end{equation}
with the same boundary conditions as for~Eq.~(\ref{eq:schrodinger}). Every eigenmode (normalized to maximum unitary amplitude) is proved to satisfy the relation
\begin{equation}\label{eq:first}
|\Psi({\bf x})|\leq E~u({\bf x})
\end{equation}
everywhere in the domain. This inequality compels the eigenfunctions to be small at the local minima of $u({\bf x})$ and along the valleys of $u$ considered as a landscape. However, due to the normalization of $\Psi$ in Eq.~(\ref{eq:first}), this constraint is only effective in the regions where $u({\bf x})<1/E$. Therefore, the portions of the valleys where $u({\bf x})$ is below $1/E$ act as confining borders for the eigenstates, thus defining localization subregions. For higher energies~$E$, the constraint is progressively lifted: neighboring localization subregions merge, up to a point where they form a set that spans the entire domain, signaling the transition to delocalized states. Consequently, while the low energy states are well confined within the valleys of $u$ (i.e., the minima of $u$ in one~dimension), higher energy states can permeate through shallow valleys and extend over several neighboring regions. This picture has been mathematically demonstrated and numerically confirmed for several random potentials~\cite{Filoche2013}.

\begin{figure}[!t]
\begin{center}
\includegraphics[width=0.4\textwidth,clip]{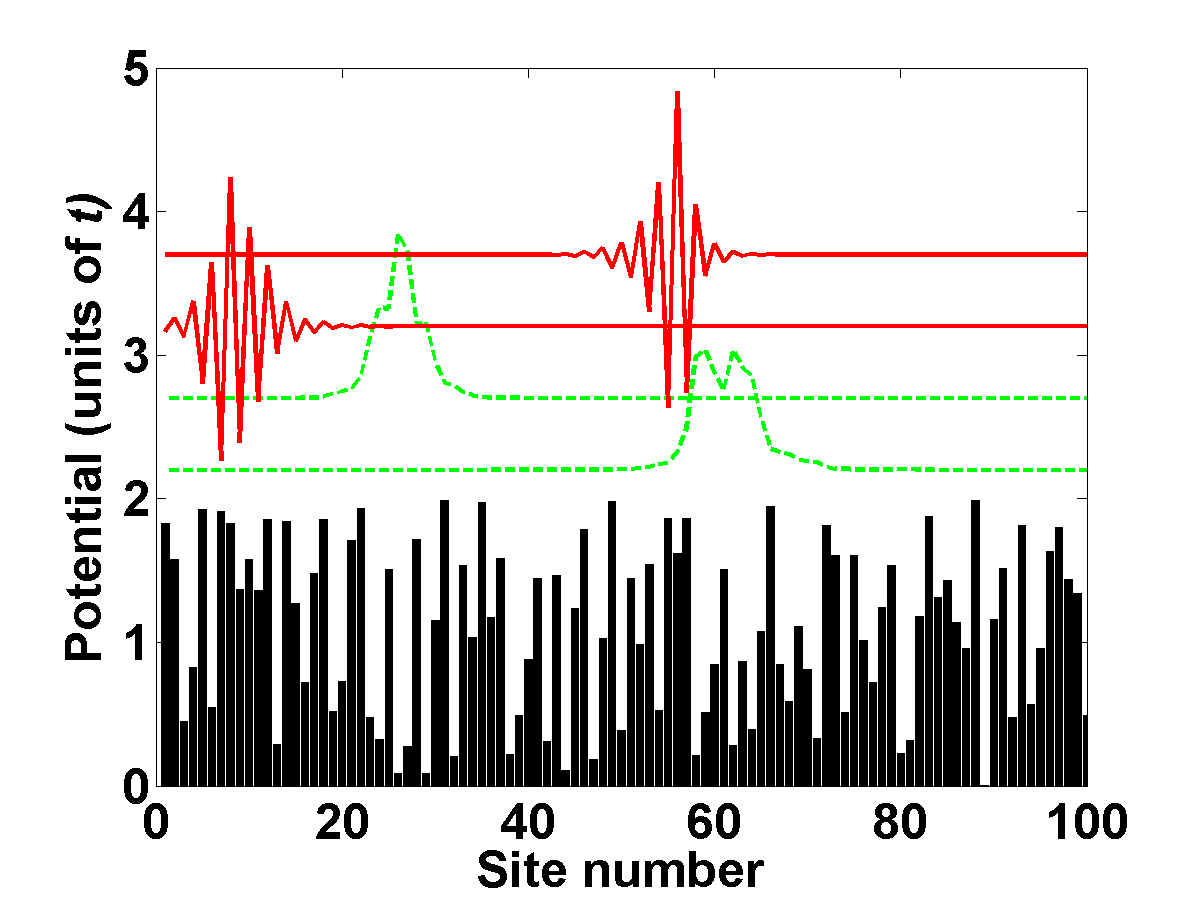}
\caption{(Color online) Bare potential landscape (filled black piecewise constant at the bottom), and the corresponding amplitudes of the two lowest energy states (dashed green lines) and of the two highest energy states (red solid lines). For better visibility, the eigenstates are vertically shifted. The eigenstates are exponentially localized in distinct chain segments. However, the bare potential landscape does not bring a clear indication of the localization subregions.}
\label{fig:potential}
\end{center}
\end{figure}

In the following, we consider the quantum mechanical tight-binding problem of one particle restricted to move along a discrete open chain with first-neighbor hopping amplitude~$t$, random on-site potentials $V_i$, and unitary lattice spacing $a=1$. For an eigenmode of energy $E$ written as a linear superposition of Wannier local orbitals $\ket{i}$, the components $(\psi_i)$ satisfy the following equation which is the discrete equivalent of the time-independent Schr\"odinger equation 
\begin{equation}
- t~[\psi_{i-1}+\psi_{i+1}] + V_i~\psi_i = E~\psi_i ~,
\label{eq:hamil}
\end{equation}
where $i \in \{1,...,L\}$ denote the chain sites and homogeneous boundary conditions are assumed, i.e., $\psi_0=\psi_{L+1}=0$. The $L$~eigenmodes can be directly obtained using standard numerical diagonalization techniques. For an on-site potential ranging from $V_{min}$ to $V_{max}$, the spectrum of possible eigen-energies is restricted to the interval $[V_{min}-2t, V_{max}+2t]$. In what follows we use energy units of $t=1$ without any loss of generality. Fig.~\ref{fig:potential} displays the two lowest and two highest energy eigenstates in a finite chain with $L=100$ sites for a realization of the random potential following a uniform distribution in the interval $[2,4]$. Note that both low and high energy states are rather localized but there is no direct indication of where to find their localization region in the original potential landscape. While both low and high energy states present exponentially localized envelope functions, they differ by an overall phase factor. This feature is reminiscent of the actual harmonic eigenstates in the limit of vanishing disorder. While the low energy states have typically a long wavelength ($k=2\pi/\lambda\rightarrow 0$), the high energy ones have wavelengths of the order of twice the lattice spacing ($k=2\pi/\lambda\rightarrow \pi/a$).

To unveil the hidden landscape confining the energy eigenstates, one has to solve the corresponding Dirichlet problem associated with the tight-binding Hamiltonian of Eq.~(\ref{eq:hamil}). A rigorous proof of Eq.~(\ref{eq:first}) satisfied by the energy eigenfunctions in the discrete case is provided in Supplemental Material~\cite{SM}. The appropriate form of the discrete Dirichlet problem satisfied by the hidden landscape becomes
\begin{equation}
-t~(u_{i+1} +u_{i-1} -2u_{i}) + (V_i-2t)~u_i = 1 ~, \qquad u_0 = u_{L+1} = 0
\label{eq:hidden1}
\end{equation}
which implies that the random potential has to be restricted to values $V_i > 2t$ to guarantee that the effective potential in the Dirichlet problem is positive everywhere.

\begin{figure}[!t]
\begin{center}
{\includegraphics[width=0.4\textwidth,clip]{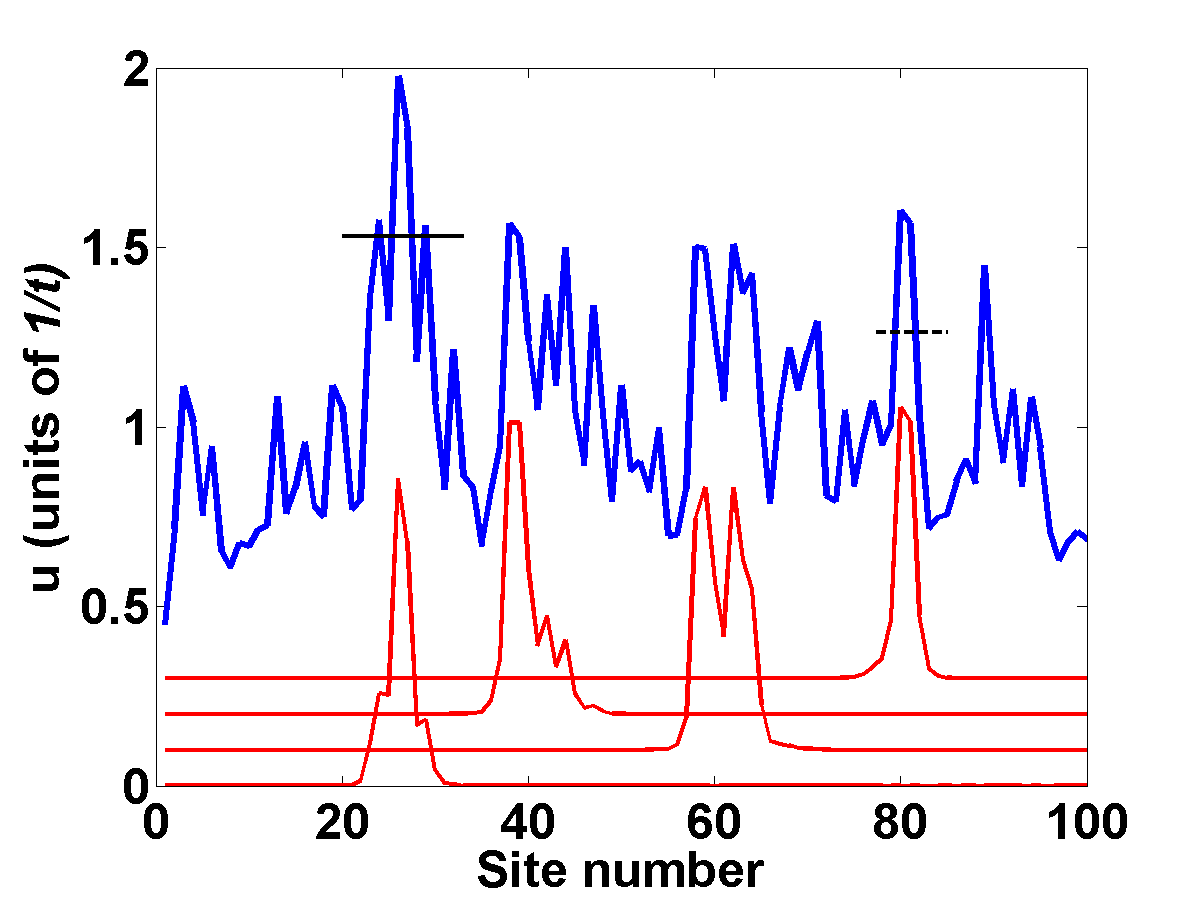}}
\caption{(Color online) Localization near the bottom of the band: the landscape $u_i$ (top black line) is plotted together with the probability distribution $|\Psi_i|^2$ of the 4 lowest energy eigenstates (bottom red lines) under the same bare potential depicted in Fig.~\ref{fig:potential}. Two horizontal line segments indicate the values of $1/E$ for the lowest energy state (left solid line) and for the 4th state (right dashed line) in the subregion where these states are localized. Note that the low energy states are trapped between the minima of $u_i$ which fulfill the confinement condition $u_i < 1/E$.}
\label{fig:landscape}
\end{center}
\end{figure}

In Fig.~\ref{fig:landscape} we plot the localization landscape~$u$ (top curve) corresponding to the random potential displayed in Fig.~\ref{fig:potential}. Below this landscape are plotted the probability amplitudes for the 4~lowest energy eigenstates (bottom curves). As predicted by the theory, the profile and location of the lowest energy states are clearly identifiable in the landscape: these states are located around the most prominent maxima of $u$ and confined by the deepest minima near each subregion. In the continuous limit where the lattice parameter goes to zero, the above equation resembles a classical Schr\"odinger equation with uniform right-hand side, and one recovers the localization of quantum states of a continuous Hamiltonian.

\begin{figure}[!t]
\begin{center}
{\includegraphics[width=0.4\textwidth,clip]{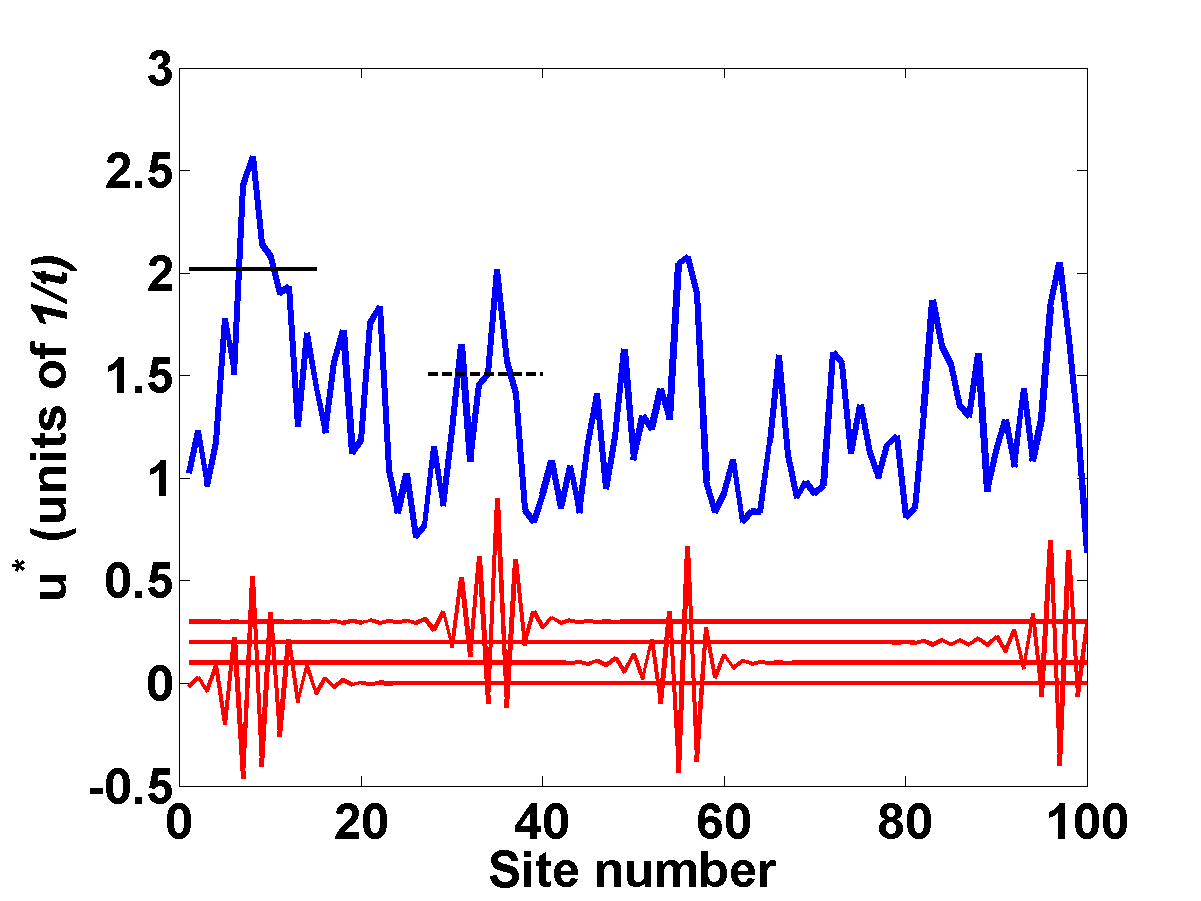}}
\caption{(Color online) Localization near the top of the band: the dual landscape $u^*$ is plotted together with the 4~eigenstates of highest energy corresponding to the same bare potential as depicted in Fig.~\ref{fig:potential}. Two horizontal line segments indicate the values of $1/(V_{\rm shift}-E)$ for the highest energy state (solid line) and for the 4$^{th}$ state from the top of the band (dashed line). Here, $V_{\rm shift} = 6$. The high energy states are localized in subregions of $u^*$ close to its most prominent peaks, and are confined between the minima of $u^*$ that fulfill the confining condition $u^*<1/(6-E)$.}
\label{fig:dual}
\end{center}
\end{figure}

Not only the theory of the localization landscape enables us to predict the occurrence of localized modes at low energy in the discrete potential, but it also explains the strong localization observed for high energy states oscillating at a scale close to the lattice parameter (corresponding to the top of the band for a periodic potential, see~Fig.~\ref{fig:dual}). To this end, one has to examine the behavior of the envelope~$\varphi$ of an eigenmode $\psi$ whose wave vector~$k = \pi/a$ is close to the top of the energy band. This envelope is defined by $\psi_i = e^{j k x_i }\varphi_i$, with $x_i = a \times i = i$ being the abscissa of site~$i$. In other words, $\varphi$ is obtained by removing the fast oscillating contribution to the eigenmode. For $k=\pi/a$, $\varphi$ obeys~\cite{SM}:
\begin{equation}
t~\left(\varphi_{i+1}+\varphi_{i-1}\right) + V_i~\varphi_i = E_i~\varphi_i
\end{equation}
One observes here two symmetry properties of the tight-binding model. First, the symmetry related to a sign change in the hopping amplitude~$t$: this symmetry reverses the energy band. The low (resp. high) energy states in a chain with hopping amplitude~$t$ become the high (resp. low) energy states when the hopping amplitude is replaced by $t \rightarrow -t$. Also, they acquire an overall phase of~$\pi/a$. As a result, the rapidly oscillating high-energy modes in a chain with hopping amplitude~$t$ are transformed into low-energy states with slowly varying envelopes in a chain with hopping amplitude~$-t$. Reversing the sign of the hopping amplitude is equivalent to reversing the signs of the original random potential landscape and of the corresponding eigen-energies. In order to avoid negative values resulting from this sign change, a global shift $V_{\rm shift}$ has to be applied to the reversed potential, which has again no consequence on the localization properties. The envelope function $\varphi$ obeys then the following Schr\"odinger-type equation:
\begin{align}
- t~\left(\varphi_{i+1}+\varphi_{i-1} - 2 \varphi_i\right) + (V_{\rm shift} - V_i -2t)~\varphi_i \nonumber\\ = (V_{\rm shift} - E_i)~\varphi_i
\end{align}
Therefore, the appropriate dual Dirichlet problem that provides the confinement landscape for the high energy states takes the form
\begin{equation}
-t~(u^*_{i+1} +u^*_{i-1} -2u^*_{i}) + (V_{\rm shift} - V_i - 2t)~u^*_i = 1 ~,
\label{eq:hidden2}
\end{equation}
where $V_{\rm shift}$ is a constant chosen such that $V_{shift} - V_i - 2t > 0$ everywhere. To keep the potential in the dual Dirichlet problem in the same range as the one in the original Dirichlet problem, the global shift has to be: $V_{\rm shift} = V_{\rm min} + V_{\rm max}$.

In Fig.~\ref{fig:dual} we show the resulting dual landscape (top curve) for the same random potential displayed in Fig.~\ref{fig:potential}, together with the 4~highest energy states (bottom curves). Although these states present spatial oscillations at the scale of the lattice parameter, the dual landscape $u^*$ clearly signals the subregions of localization close to its most prominent maxima. Also, the confinement strength decreases as one departs from the top of the band, the confinement condition at $u_i < 1/E$ being replaced by $u^*_i < 1/(V_{\rm shift}-E)$.

We now show that these landscapes $u$ and $u^*$ can allow us to compute an estimate of the average localization length of the eigenstates around any given energy. This estimate will then be compared to the participation ratio of each mode, a widely used measurement of the localization length~\cite{Thouless1974,Prelovsek1978}, defined as
\begin{equation}
\label{eq:Pk}
P =  \left(\sum_i|\psi_i|^4\right)^{-1}
\end{equation}
where $\psi_i$ are the coefficients of the normalized eigenmode expanded in the local Wannier basis states ($|\Psi\rangle = \sum_i\psi_i|i\rangle$). This ratio is usually understood as a measure of the number of sites on which the particle probability distribution is concentrated. It equals the total number of lattice sites for a uniformly distributed state while it is of the order of the localization length for exponentially localized states. For the one-dimensional tight-binding model with uniform potential, all harmonic eigenstates have an identical participation ratio equal to $2L/3$ where $L$ is the chain size.
 
\begin{figure}[!t]
\begin{center}
{\includegraphics[width=0.35\textwidth,clip]{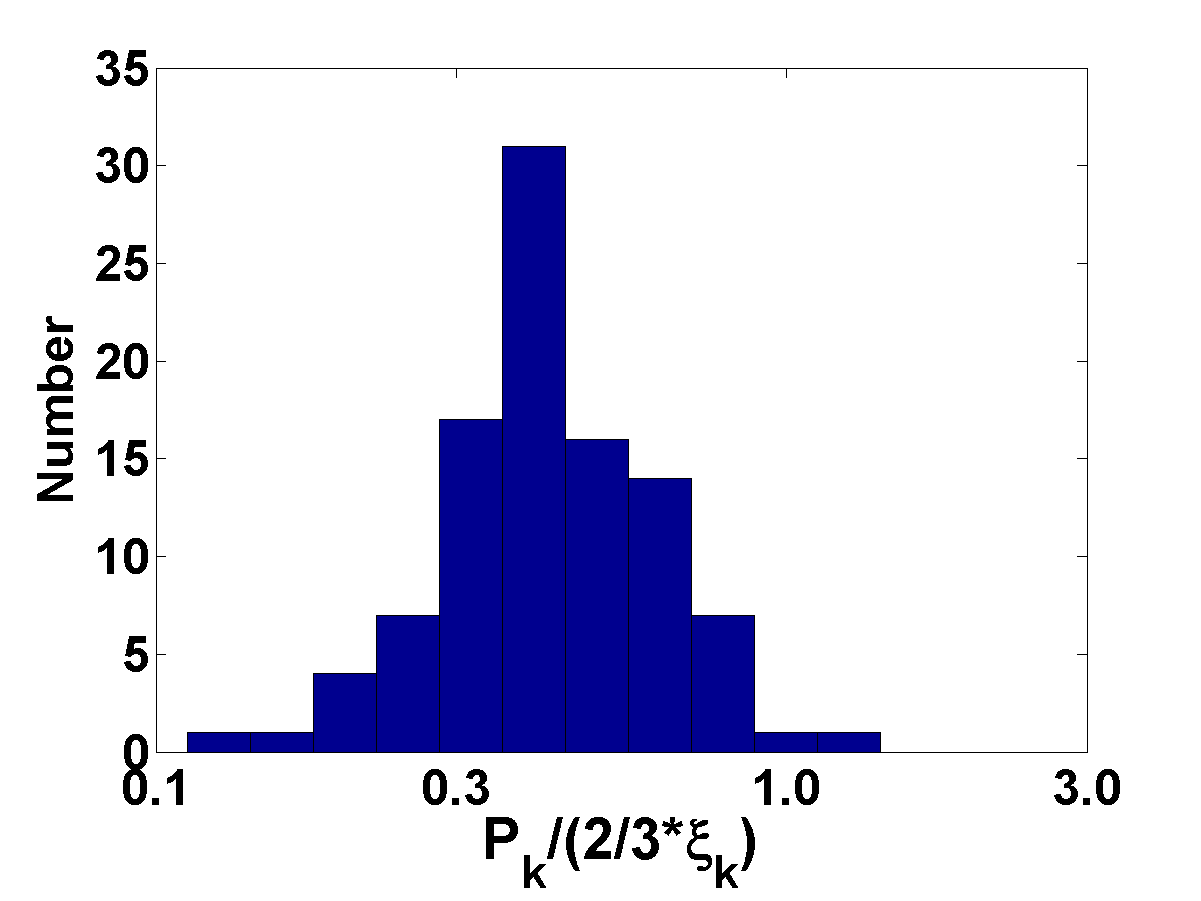}}
\caption{Histogram in log scale of the ratio $P_k/(2/3 \xi_k)$ for the bottom 50 and top 50 eigenstates in a chain of $L=200$ sites and a random potential $V$ whose values range in the interval $[2;10]$. The participation ratio of an eigenstate $\psi_k$ is defined in Eq.~(\ref{eq:Pk}). The size of its confining subregion $\xi_k$ is determined using the landscape $u$ for the eigenstates at the bottom of the band, and using the dual landscape $u^*$ at the top of the band. This histogram shows that $P_k$ is always of the same order as $\xi_k$ for localized eigenstates.}
\label{fig:hist}
\end{center}
\end{figure}

\begin{figure}[!t]
\begin{center}
{\includegraphics[width=0.235\textwidth,clip]{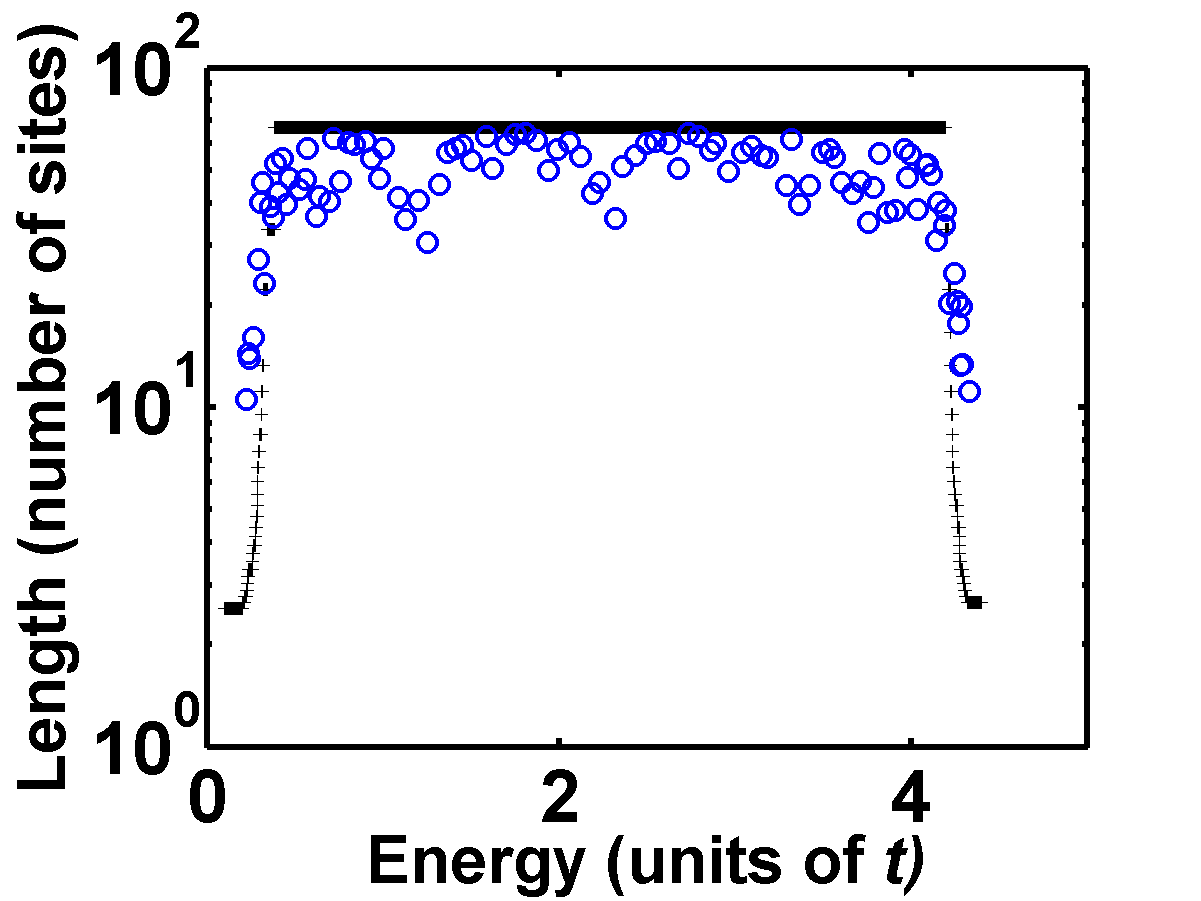}}
{\includegraphics[width=0.235\textwidth,clip]{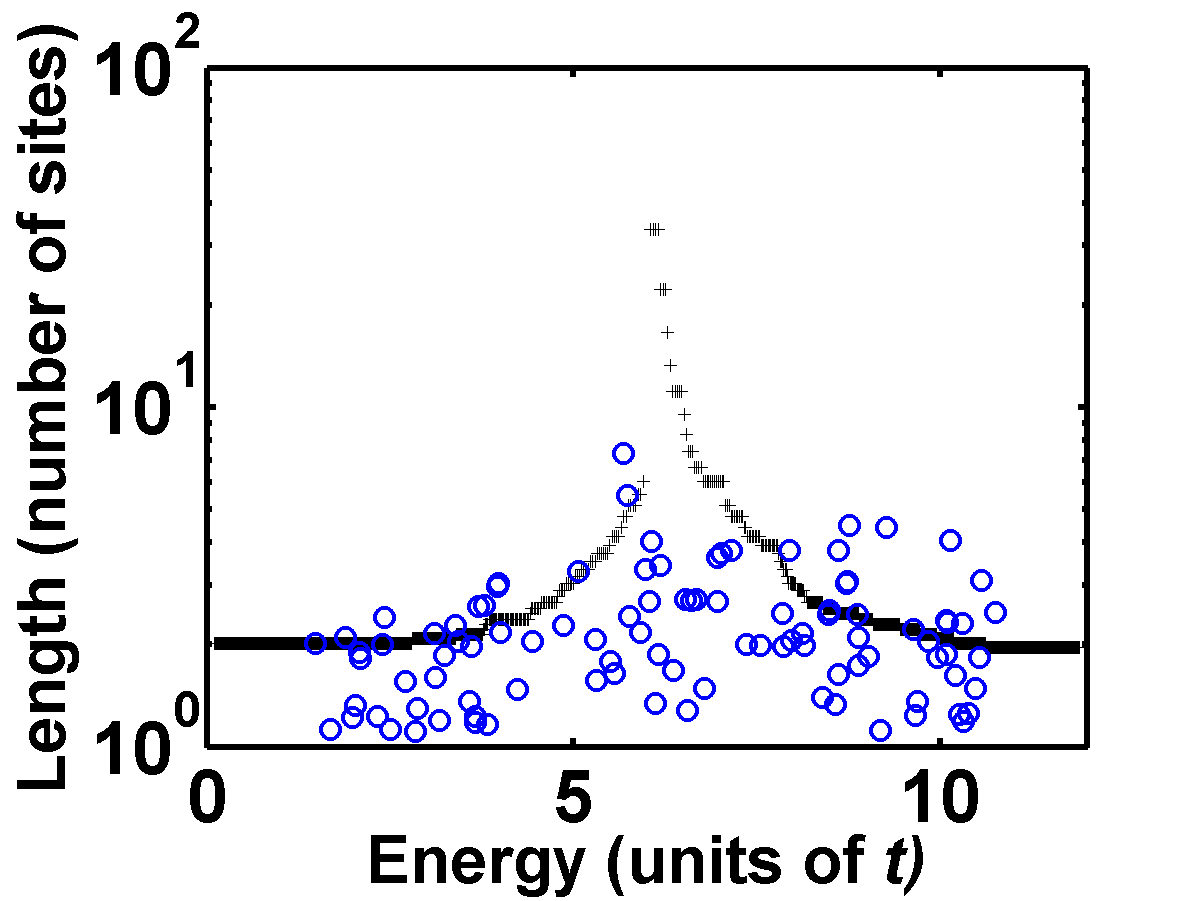}}
\caption{Participation ratio (circles) and $\delta(E)$ (crosses), the latter corresponding to 2/3 $\times$ the average size of the confining subregions for eigenfunctions of energy smaller than $E$ (see below), in two different cases. (left) Weak disorder regime ($V=[2;2.5]$). Notice that the average size of the confining subregions closely follows the main trends of the participation number in the whole energy spectrum; (right) Strong disorder regime ($V=[2;10]$). The length $\delta(E)$ computed using the two potentials $u$ and $u^*$ capture the strong localization of the eigenstates.}
\label{fig:spectrum}
\end{center}
\end{figure}

To pinpoint the confining sub-region -- defined by the landscape~$u$) -- associated with an eigenstate of low energy~$E_k$, we first locate the chain site $i_{max}$ at which this specific mode has its largest amplitude. We then define the size of the localization subregion $\xi_k$ containing this specific eigenstate as the length $(j-i)~a$ of the smallest interval $[i,j]$ containing $i_ {max}$ (i.e. $i < i_{max} < j$) such that $u_i$ and $u_j$ are local minima of $u$ that are both smaller than $1/E_k$. In other words, $i$ and $j$ are the nearest ``valleys'' of the landscape~$u$ surrounding this eigenstate. For high-energy modes, a similar procedure is employed using the dual landscape $u^*$ that is compared to the inverse of the energy distance to the band top as the confinement strength criterion.

In Fig.~\ref{fig:hist} we evidence the very strong correlation between the participation ratio and the size of the confining subregions determined by the above procedure by plotting the histogram of the ratio $P_k / (2/3 \xi_k)$ for the 50 states of lowest energy and the 50 states of highest energy in a 200-site chain (i.e., half of the states, their total number being 200). Note that $P_k / (2/3 \xi_k)$ can take in theory any value ranging from $1/(2/3 L)$ to $L/(4/3)$, where $L=200$ is the chain length. Although, the participation ratio of an individual eigenstate may vary from a few sites to the entire length of the chain, both quantities appear to be always of the same order of magnitude.

This strong correlation between the participation ratio and the size of the confining subregions provides a key to introducing an even simpler estimate of the localization length, called here $\delta(E)$ which depends only on the energy value~$E$. It is obtained by averaging all distances between minima of $u$ which are below $1/E$, and then multiplying the quantity obtained by 2/3. As reinforced above, the minima of the landscape $u$ act as confinement points in 1D. However, for a specific low-energy eigenmode of energy $E$, the confinement is effective at a given minimum only when the value of $u$ (resp. $u^*$) at this point is below $1/E$ (resp. $1/(V_{shift}-E)$). Therefore $\delta(E)$ can be understood as averaging the sizes $\xi$ of the subregions found below $1/E$, independently of their location in the system. According to the localization theory developed in~\cite{Filoche2012}, this distance can be used as a rough estimate of the average localization length of all eigenmodes with energy in a small energy window around~$E$.

In Fig.~\ref{fig:spectrum}, we superimpose the plots of the participation ratio for each eigenstate (circles) and the average size $\delta(E)$ of the localization subregions (line). For the first half of eigenmodes (corresponding to the first half of the band), $\delta(E)$ is computed using the landscape $u$, whereas the dual landscape $u^*$ is used for the second half of the energy band. Two representative cases of disorder, weak and strong (resp. Fig~\ref{fig:spectrum}a and Fig~\ref{fig:spectrum}b), are illustrated. In the case of weak disorder, $\delta(E)$ reproduces the behavior of the average participation ratio over the entire energy band, correctly predicting the location of the pseudo-mobility edges separating the well localized from the delocalized states. These delocalized states have an harmonic-like form with random phase changes. Consequently, their participation ratios fluctuate around $2L/3$. Accordingly, $\delta(E)$ reaches a plateau at $2L/3$ in the energy range corresponding to effectively delocalized states signaling that the dual landscapes have no minima below the threshold level in this energy range.

In the regime of strong disorder all states become well localized with no pseudo-mobility edges within the band of allowed energies. The average size $\delta(E)$ also captures such strong localization, although near the center of the band it signals a weaker localization. This behavior reflects the fact that the continuous Laplacian operator does not properly describe all features of the discrete tight-binding Hamiltonian near the band center.

In summary, we have shown here that the localization of one-particle eigenstates satisfying a discrete time-independent Schr\"odinger equation is in reality governed by a pair of dual landscapes, $u$ and $u^*$, a priori invisible to the naked eye, respectively acting on the regimes of low and high energies. Using the one-dimensional tight-binding Hamiltonian with diagonal disorder as a prototype model, we demonstrated that the appropriate Dirichlet problem whose solution unveils the landscape of localization has distinct forms near the bottom and the top of the energy band. Using the symmetries of the tight-binding Hamiltonian, the landscape confining the high-energy states is found to be the solution of an alternate Dirichlet problem with a new potential. These dual landscapes signal the localization subregions in both energy regimes, a task that had not been successfully achieved in prior studies aiming to provide a geometrical interpretation for the Anderson localization.

The distinct confinement strengths of the hidden landscapes were used to introduce a new measure of the localization length. Specifically, the energy eigenmodes are confined between minima of the hidden landscape that are below a threshold level given by the inverse of the energy distance to the band edge. We showed that the average size of the confining subregions $\delta (E)$, despite its approximate nature, captures well the main dependence of the localization length within the range of allowed energies, especially in the regime of weak disorder at which $\delta (E)$ clearly signals the location of the pseudo-mobility edges separating well localized from effectively delocalized states.

The present scenario opens a totally new perspective to the geometric interpretation of Anderson localization in discrete lattices based on the hidden landscapes $u$ and $u^*$. Specifically, in 2 or 3 dimensions, one can extrapolate that $u$ will remain as the low energy landscape while $u^*$ has to be replaced by a collection of landscapes, each corresponding to one boundary of the first Brillouin zone of the lattice. This approach also raises a number of new questions that remain to be addressed. Could it be possible to unify the present description based on several distinct landscapes into a wavevector-dependent landscape scenario valid for the whole energy band? Can these landscapes signal resonant delocalized states that are usually depicted by discrete tight-binding models with correlated disorder?~\cite{Dunlap1990,Wu1991,Hilke1997,deMoura2003} In particular, inter-particle interactions are known to influence the localization properties~\cite{Srinivasan2003,Dias2007,Song2008,Henseler2008a,Henseler2008b}. In this context, it would be very interesting to assess how these interactions can distort the landscapes. New analytical and numerical efforts along these directions will certainly contribute to unveil a new geometrical picture of the Anderson localization in all discrete lattices.


This work was partially funded by the Brazilian Research Agencies CAPES, CNPq, FINEP  and FAPEAL. The sabbatical stay of Marcelo Lyra at Physique de la Mati\`ere Condens\'e Lab was funded through the Grant ATLAS from the Triangle de la Physique. Marcel Filoche is partially supported by a PEPS-PTI Grant from CNRS. Svitlana Mayboroda is partially supported by the Alfred P. Sloan Fellowship, the NSF CAREER Award DMS 1056004, the NSF MRSEC Seed Grant, and the NSF INSPIRE Grant.

\bibliography{localization}

\end{document}


\section{Control inequality for a discrete Schr\"odinger operator}

The present supplement provides mathematical proofs of the main control inequalities. The arguments are fairly straightforward, but it is important that they are proved directly in the discrete scenario (rather than appealing to approximation by a continuous model), so for completeness we present here a full version. 

Let us denote by $\Delta_D$ the discrete Laplacian, that is, for $u:=(u_1,...,u_L)^{\perp}$ (colon vector) we write
\begin{eqnarray}\label{defLap}
\left(\Delta_D u\right)_i := u_{i+1} + u_{i-1} - 2 u_i 
\end{eqnarray}
Given a potential $V_i$ defined at each site, we consider the Schr\"odinger-type operator $-t\Delta_D + W$ such that
\begin{eqnarray}\label{defLap2}
\left(-t\Delta_D + W u\right)_i &:=& -t~( u_{i+1} + u_{i-1} - 2 u_i) + (V_i-2t) ~u_i \nonumber \\ 
&~=&-t~(u_{i+1}+u_{i-1}-2 u_i) + W_i~u_i,
\end{eqnarray}
where $t>0$ and $W=(W_1,...,W_L)^{\perp}$ is such that $V_i\geq2t$ for all $i$ and hence, $W_i\geq 0$ for all $i$ (slightly abusing the notation, we write $W\geq 0$). Clearly, the operator  $-t\Delta_D+W$ can be identified with multiplication from the left by a matrix $A$ with values $A_{ii}=2t+W_i$, $i=1,..., L$, on the main diagonal, $W_{i, i-1}=-t$, $i=2, ..., L$, $W_{i+1, i}=-t$, $i=1, ..., L-1$ on lower and upper diagonal, and 0 otherwise. 

\begin{lemma}[Maximum principle]\label{lMaxPr} Let $t>0$ and $W\geq 0$ as above.
If $\left((-t\Delta_D+W) u\right)_i\geq 0$, $i=1,...,L$, and $u(0)=u(L+1)=0$, then $u_i\geq 0$ for all $i=1,...,L$. Moreover, if there exists an $i_0$ such that $\left((-t\Delta_D+W) u\right)_{i_0}> 0$ then $u_i>0$ for all $i=1,..., L$.
\end{lemma}

\bp We prove by contradiction. Assume that there is a minimum ``inside" the domain, that is, there exists $i_0$ such that $u_{i_0}\leq u_{i_0+1}$ and $u_{i_0}\leq u_{i_0-1}$. Then $u_{i+1}+u_{i-1}-2 u_i\geq 0$. But $\left((-t\Delta_D+W) u\right)_i\geq 0$, hence, $t(u_{i+1}+u_{i-1}-2 u_i)\leq W_iu_i$. Hence, $u_i\geq 0$, but we assumed that it was a minimum and hence, that it was below the boundary values equal to zero. This is a contradiction. We conclude that the minimum could not be ``inside" the domain and hence, that all interior values of $u$ are non-negative. 

In order to show strict positivity, it is enough to demonstrate that if $u_{i_0}=0$ at some $i_0$, then $\left((-t\Delta_D+W) u\right)_i=0$ for all $i$. To this end, assume that $u_{i_0}=0$ at some $1\leq i_0\leq L$. Since it cannot be a local minimum, the values of $u_{i_0-1}, u_{i_0+1}$ must be smaller or equal than $u_{i_0}=0$. Since they cannot be below zero, we have $u_{i_0-1}, u_{i_0+1}=0$. Continuing in this fashion, we conclude that $u_0\equiv 0$ and hence, $\left((-t\Delta_D+W) u\right)_i=0$ for all $i$, as desired.
\ep

\begin{lemma}[Positivity]\label{lPos}
For the matrix $A$ associated with the operator $-t\Delta_D+W$ as above, the inverse exists and every entry of the inverse is strictly positive. 
\end{lemma}

This is an analogue of the positivity of the Green function.

\bp Let $\left((-t\Delta_D+W) u\right)_i=f_i$, $i=1,...,L$, $u_0=u_{L+1}=0$. In matrix notation, $A\vec u=\vec f$. Then $u_j=(A^{-1} f)_j=\sum_{k=1}^L (A^{-1})_{jk} f_k$. If we take $\vec f$ as a vector with 0 entries except for 1 in the $l$-th place, then $u_j=(A^{-1})_{jl}$. By (the second statement of) Lemma~\ref{lMaxPr}, all $u_j$, $j=1,...,L$ must be strictly positive, as desired. \ep

\begin{lemma}[First inequality in the discrete case]\label{lIneq1}
Let 
$$\left((-t\Delta_D+W) \psi \right)_j=\lambda \psi_j, \quad j=1,...,L, \quad \psi_0=\psi_{L+1}=0,$$
where the operator is defined as in \eqref{defLap2} with $t>0$ and it is assumed that $W=V-2t\geq 0$. Then
$$\frac{|\psi_j|}{\max_k|\psi_k|}\leq \lambda u_j, \quad\mbox{for all } j=1,...,L,$$
where $u$ solves
$$\left((-t\Delta_D+W) u\right)_j=1, \quad j=1,...,L, \quad u_0=u_{L+1}=0.$$
\end{lemma} 

\bp This is a simple consequence of the Lemmas above. Indeed, using the positivity established in Lemma~\ref{lPos}, 
\begin{eqnarray}
\psi_j=(\lambda A^{-1}\psi)_j&=&
 \lambda\sum_{k=1}^{L} (A^{-1})_{jk}\psi_k\leq \lambda \max_k|\psi_k| \sum_{k=1}^L(A^{-1})_{jk} \nonumber \\
~&=&\lambda \max_k |\psi_k| u_j. \nonumber
\end{eqnarray}\ep

Lemma~\ref{lIneq1}, applied, as above, with $W=V-2t$, furnishes inequality (3) in the main manuscript of the paper for the discrete model, and respectively, treats the lower energy modes. To address the higher order ones, we apply the transform $\psi_j=e^{i\alpha j}\varphi_j$, $j=1,...,L$, $\alpha\in\RR$. Note the change of notation: from now on, $i$ is the imaginary unit. Then the eigenfunctions of $(-t\Delta_D+W)=(-t\Delta_D+V-2t)$ as above, denoted by $\psi$, are transformed into the ``envelope" functions $\varphi$ satisfying the following equation:
$$ -t (e^{i\alpha}\varphi_{j+1}+e^{-i\alpha}\varphi_{j-1})+V_j \varphi_j=\lambda \varphi_j.$$
The choice $\alpha=\pi$, optimal for studying the eigenfunctions at the high end of the band gap, yields 
$$ t (\varphi_{j+1}+\varphi_{j-1})+V_j \varphi_j=\lambda \varphi_j,$$
or, as discussed in the paper, 
$$ -t (\varphi_{j+1}+\varphi_{j-1}-2\varphi_j)+(V_{shift}-2t-V_j) \varphi_j=(V_{shift}-\lambda) \varphi_j,$$
where $V_{shift}$ is chosen to ensure that $V_{shift}-2t-V_j\geq 0$. 

Zero Dirichlet boundary conditions on $\psi$ trivially yield zero Dirichlet boundary conditions for $\varphi$, and hence, Lemma~\ref{lIneq1} yields the corollary. 
\begin{corollary}[First inequality for the dual landscape]\label{cIneq1-dual}
Let 
$$\left((-t\Delta_D+W) \psi \right)_j=\lambda \psi_j, \quad j=1,...,L, \quad \psi_0=\psi_{L+1}=0,$$
where the operator is defined as in \eqref{defLap2} with $t>0$.
Let furthermore $\varphi_j$ denote the envelopes of $\psi_j$, defined via $\psi_j=e^{i\alpha j}\varphi_j$, $j=1,...,L$.
Then for any choice of $V_{shift}$ such that $V_{shift}-2t-V\geq 0$, we have
$$\frac{|\varphi_i|}{\max_j|\varphi_j|}\leq (V_{shift}-\lambda) u^*_i, \quad\mbox{for all } i=1,...,L,$$
where $u$ solves
$$\left((-t\Delta_D+W^*) u^*\right)_i=1, \quad i=1,...,L, \quad u_0=u_{L+1}=0,$$
with $W^*=V_{shift}-2t-V\geq 0$.
\end{corollary}